\documentstyle[12pt,aaspp4,epsfig]{article}
\begin{document}

\title{Gravitational Collapse in Turbulent Molecular Clouds. II.
       Magnetohydrodynamical Turbulence}

\author{Fabian Heitsch\altaffilmark{1}}
\author{Mordecai-Mark Mac Low\altaffilmark{2}}
\author{Ralf S. Klessen\altaffilmark{3}}
\altaffiltext{1}{Max-Planck-Institut f\"ur Astronomie, K\"onigstuhl 17, D-69117
Heidelberg, Germany; E-mail: heitsch@mpia-hd.mpg.de}
\altaffiltext{2}{Department of Astrophysics, American Museum of
Natural History, Central Park West at 79th Street, New York, New York
10024-5192, USA; E-mail: mordecai@amnh.org}
\altaffiltext{3}{Sterrewacht Leiden, The Netherlands; E-mail:
klessen@strw.leidenuniv.nl} 


\begin{abstract}

Hydrodynamic supersonic turbulence can only prevent local
gravitational collapse if the turbulence is driven on scales smaller
than the local Jeans lengths in the densest regions, a very severe
requirement (Paper I). Magnetic fields have been suggested
to support molecular clouds either magnetostatically or via
magnetohydrodynamic (MHD) waves. Whereas the first mechanism would
form sheet-like clouds, the second mechanism not only could exert a
pressure onto the gas counteracting the gravitational forces, but
could lead to a transfer of turbulent kinetic energy down to smaller
spatial scales via MHD wave interactions. This turbulent magnetic
cascade might provide sufficient energy at small scales to halt local
collapse.

We test this hypothesis with MHD simulations at resolutions up to
$256^3$ zones, done with ZEUS-3D.  We first derive a resolution
criterion for self-gravitating, magnetized gas: in order to prevent
collapse of magnetostatically supported regions due to
numerical diffusion, the minimum Jeans length must be resolved by four
zones.  Resolution of MHD waves increases this requirement to roughly
six zones.  We then find that magnetic fields cannot prevent local
collapse unless they provide magnetostatic support.  Weaker
magnetic fields do somewhat delay collapse and cause it to occur more
uniformly across the supported region in comparison to the
hydrodynamical case.  However, they still cannot prevent local
collapse for much longer than a global free-fall time.

\end{abstract}


\clearpage

\section{Introduction}

All star formation takes place in molecular clouds. However, the star
formation rate in these clouds is surprisingly low. From the
Jeans argument one would expect that they should collapse within their
free fall time
\begin{equation}
t_{\rm ff} = \sqrt{\frac{3\pi}{32 G \bar\rho}} \approx (1.06 \times 10^6
\mbox{ yr}) \left(\frac{n}{10^3 \mbox{ cm}^{-3}}\right)^{-1/2},
\label{equ:tff}
\end{equation}
where $\bar\rho$ is the mean mass density of the cloud, $G$ the gravitational 
constant and $n=\bar\rho/\mu$ the number density, with $\mu=2.36 m_H$. 
 A catastrophic collapse of a giant molecular 
cloud on this timescale would yield a
single starburst event.  However, molecular clouds have classically
been thought to survive without global collapse for much longer than
their free-fall time $t_{\rm ff}$ (Blitz \& Shu 1980).  Moreover, stars
are not usually observed in nearby star-forming regions to form in
such a catastrophic collapse. Instead, they form in localized regions
dispersed through an {\em apparently} stable cloud.

Observations of spectral line widths in molecular clouds show that the
gas moves at speeds exceeding the thermal velocities by up to an order
of magnitude (Williams et al.\ 1999).  These supersonic motions seem
not to be ordered, so that turbulent support models have often been
suggested, with the turbulence giving rise to an effectively isotropic
turbulent pressure counteracting the gravitational forces.  

However, such models have two problems.  First, simulations of
supersonic, compressible turbulence show that it typically decays in a
time less than the cloud's free-fall time $t_{\rm ff}$ (Gammie \& Ostriker
1996, Mac Low et al. 1998, Mac Low 1999).  So, in order to be really 
able to support the cloud, the turbulence would have to be constantly 
driven. Second, although hydrodynamical turbulence can prevent {\em
global} collapse, it can never completely prevent {\em local} collapse
except with unrealistically short driving length scale $\lambda_D$
(Klessen, Heitsch \& Mac Low 2000; hereafter Paper I). The efficiency 
of local collapse depends on the wavelength and on the
strength of the driving source. Long wavelength driving
or no driving at all results in efficient, coherent star formation,
with most collapsed regions forming near each other (Klessen \& Burkert 2000).  
Strong, short wavelength driving, on the other hand, results in inefficient,
incoherent star formation, with isolated collapsed regions randomly 
distributed throughout the cloud.

The model of molecular clouds being supported by turbulence has been
widely discussed and investigated, as reviewed in Paper I and
V\'azquez-Semadeni et al. (2000).  Recently, Ballesteros-Paredes et
al.~(1999), Hennebelle \& P\'erault (1999), and Elmegreen (2000) have
suggested that molecular clouds might not have to be supported for
these long time scales at all, but might be transient features caused
by colliding flows in the interstellar medium.  This would solve very
naturally not only the problem of cloud support, but also, according
to Ballesteros-Paredes et al.~(1999), the pronounced lack of 5--10
million year old post-T~Tauri-stars directly associated with
star-forming molecular clouds.

Magnetic fields might alter the dynamical state of a molecular cloud
sufficiently to prevent gravitationally unstable regions from
collapsing (McKee 1999).  They have been hypothesized to support
molecular clouds either magnetostatically or dynamically through MHD
waves.  

Mouschovias \& Spitzer (1976) derived an expression for the
critical mass-to-flux ratio in the center of a cloud for magnetostatic
support. Assuming ideal MHD, a self-gravitating cloud of
mass $M$ permeated by a uniform flux
$\Phi$ is stable if the mass-to-flux ratio
\begin{equation}
  \frac{M}{\Phi} <
  \left(\frac{M}{\Phi}\right)_{cr}\equiv\frac{c_\Phi}{\sqrt{G}}.
  \label{equ:magstatsup}
\end{equation}
with $c_\Phi$ depending on the geometry and the field and density
distribution of the cloud.  A cloud is termed {\em subcritical} if it
is magnetostatically stable and {\em supercritical} if it is not.
Mouschovias \& Spitzer (1976) determined that $c_\Phi = 0.13$ for
their spherical cloud. Assuming a constant mass-to-flux ratio in a
region results in $c_\Phi = 1/(2\pi) \simeq 0.16$ (Nakano \& Nakamura
1978). Without any other mechanism of support such as turbulence
acting along the field lines, a magnetostatically supported cloud will
collapse to a sheet which then will be supported against further
collapse. Fiege \& Pudritz (1999) discussed a sophisticated version of
this magnetostatic support mechanism, in which poloidal and toroidal
fields aligned in the right configuration could prevent a cloud
filament from fragmenting and collapsing.

Investigation of the second alternative, support by MHD waves,
concentrates mostly on the effect of Alfv\'{e}n waves, as they (1) are
not as subject to damping as magnetosonic waves and (2) can exert a
force {\em along} the mean field, as shown by Dewar (1970) and Shu et
al.~(1987). This is because Alfv\'{e}n waves are {\em transverse}
waves, so they cause perturbations $\delta \mathbf{B}$ perpendicular to
the mean magnetic field $\mathbf{B}$. McKee \& Zweibel (1994) argue that
Alfv\'{e}n waves can even lead to an isotropic pressure, assuming that
the waves are neither damped nor driven. However, in order to support
a region against self-gravity, the waves would have to propagate
outwardly, rather than inwardly, which would only further compress the
cloud. Thus, as Shu et al.~(1987) comment, this mechanism requires a
negative radial gradient in wave sources in the cloud.

Most molecular clouds show evidence of magnetic fields.  However, the
discussion of their relative strength is lively.  Crutcher (1999)
summarizes all 27 available Zeeman measurements of magnetic field
strengths in molecular clouds. He concludes that (1) static magnetic
fields are not strong enough to support the observed clouds alone,
with typical ratios of the mass $M$ to the critical mass for the
observed magnetic field $M/M_{cr} \approx 2$; (2) the ratio of
thermal to magnetic pressure $\beta = P_{th}/P_{mag} \approx 0.04$;
(3) internal motions are supersonic, with a velocity dispersion
$\sigma_v \gg c_s$ but approximately equal to the Alfv\'{e}n speed,
$\sigma_v \approx v_A$; and (4) that the kinetic and magnetic energies
are roughly equal, which he interprets as 
suggesting that static magnetic fields and MHD
waves are equally important in cloud energetics.  However, McKee (1999)
remarks that (a) Crutcher's data do not address the strength of the
field on large scales (threading an entire GMC) and (b) that the data
deal with dense regions in the clouds, so that ambipolar diffusion
already might have altered the mass-to-flux ratio observed.  Moreover,
Crutcher's fourth conclusion about the role of static fields and MHD
waves is based on the assumption that the kinetic energy stems mainly
from MHD waves.  

Nakano (1998) made two further arguments against magnetostatic support 
of cloud cores.  First, magnetically subcritical
condensations cannot have column densities much higher than their
surroundings.  However, observed cloud cores have column densities
significantly higher than the mean column density of the cloud,
indicating that they are not magnetostatically supported.  Second, if
the cloud cores were magnetically supported and subcritical it would
be difficult to maintain the observed non-thermal velocity dispersions
for a significant fraction of their lifetime.  Mac Low (1999)
confirmed this by numerically determining the dissipation rate of
supersonic, magnetohydrodynamic turbulence.  He concludes that the
typical decay time constant is far less than the free-fall time of the
cloud.

Polarization measurements might give us a clue whether the fields are
well ordered or in a turbulent state.  However, up to now, most
measurements refer to the highest density regions, thus giving
information about the fields in small scale structures, but not about
scales of the whole cloud.  Hildebrand et al. (1999) present
polarization measurements of cloud cores and envelopes. They find
polarization degrees of at most $10$\%. More recent observations of
the molecular cloud filament OMC-3 by Matthews \& Wilson (1999)
suggest that the magnetic field is well ordered perpendicularly to the
filament, but with a mean polarization degree of only $4.2$\%.

V\'azquez-Semadeni et al. (1996) performed three-dimensional
simulations including self-gravity and MHD with a resolution of
$64^3$. They found that hydrodynamical and supercritically magnetized
turbulence can lead to gravitationally bound structures.  Gammie \&
Ostriker (1996) did simulations in $1\mbox{ }2/3$ dimensions, while
more recently $2.5$ dimensional models were presented by Ostriker et
al. (1999). Mac Low et al. (1998), Stone et al. 1998, Padoan \&
Nordlund (1999) and Mac Low (1999) studied decaying magnetized
turbulence and found short decay rates with as well as without
magnetic fields.

We present the first high-resolution ($256^3$ zones) simulations of
magnetized, self-gravi\-tating, driven, supersonic turbulence to test
the hypothesis that magnetic fields can contribute to the support of
molecular clouds.  The following section describes the technique and
parameters used for the simulations. In section \ref{sec:resolution}
we discuss requirements regarding the resolution needed for
simulations of self-gravitating magnetized turbulence. In section
\ref{sec:results} we present the results, and we summarize our
conclusions in section \ref{sec:conclusions}.

\section{Technique and Models\label{sec:technique_models}}

\subsection{Technique \label{subsec:technique}}

For our computations, we use ZEUS-3D, a well-tested, Eulerian,
finite-difference code (Stone \& Norman 1992a, b; Clarke 1994). It
uses second-order advection and resolves shocks employing a von Neumann
artificial viscosity.  Self-gravity is implemented via an FFT-solver
for cartesian coordinates (Burkert \& Bodenheimer 1993). The magnetic
forces are calculated via the constrained transport method (Evans \&
Hawley 1988) to ensure ${\mathbf\nabla}\cdot{\mathbf B}=0$ to machine
accuracy.  In order to stably propagate shear Alfv\'{e}n-waves, ZEUS
uses the method of characteristics (Stone \& Norman 1992b, Hawley \&
Stone 1995). This method evolves the propagation of Alfv\'{e}n-waves
as an intermediate step to compute time-advanced quantities for the
evolution of the field components themselves to ensure that signals do
not propagate upwind unphysically.

We use a three-dimensional, periodic, uniform, Cartesian grid for the
models described here.  This gives us equal resolution in all regions,
and allows us to resolve shocks and magnetic field structures well
everywhere.  On the other hand, collapsing regions cannot be followed
to scales less than a few grid zones.

We do not include ambipolar diffusion in this work, although numerical
diffusion acts on scales in our simulations that, when translated to
astronomical scales with typical parameters, correspond to the scales
on which ambipolar diffusion begins to dissipate power (see
Paper~I). Moreover, we do not use a physical resistivity, relying on
numerical dissipation at grid scale, due to averaging quantities over
zones.  The limited resolution in high-density regions can lead to
excessive numerical diffusion of mass through the magnetic fields that
must be accounted for when analyzing these or similar computations.  In 
\S~\ref{sec:resolution} we derive the appropriate resolution criterion.

\subsection{Models \& Parameters \label{subsec:models}}

We employ two sets of parameters. The first one is the same as in Paper
I and allows us to compare its results to the ones of this work.
The second parameter set enables us to determine the numerical
reliability of these results.  All parameters are given in normalized
units, where physical constants are scaled to unity and where we
consider gas cubes with mass $M \equiv 1$ and side length
$[-1.0,1.0]$. The system can be scaled to physical units using the
Jeans mass $M_J$ and length scale $\lambda_J$. In Paper I we adopt a
normalized sound speed of $c_s=0.1$, which yields $M_J=0.0156$ and
$\lambda_J=0.5$ such that the computed box contains $M=64\,M_J$ and is
$L=4\,\lambda_J$ in size. Our second parameter
set is based on a sound speed of $c_s=0.213$, which in turn gives a 
ten times larger Jeans mass $M_J=0.156$ and yields $\lambda_J=1.068$, 
so that follows $M=6.4\,M_J$ and $L=1.873\,\lambda_J$. 

We use the same driving mechanism described in Mac Low (1999). Each
time step, a fixed velocity pattern is added to the actual velocity
field thus, that the input energy rate is constant. The driving field
pattern is derived from a Gaussian random field with a given spectrum.
This allows us to
drive the cloud on selected spatial scales. The {\em rms} Mach number
for the first parameter set is ${\cal M}_{rms}=10$, for the second one
it is ${\cal M}_{rms}=5$. 

The MHD simulations start with a uniform magnetic field in the 
$z$-direction. As soon as the cube has evolved into a fully turbulent
state, gravity is switched on.  The critical field value according to
equation \ref{equ:magstatsup} is $B_{cr}=1.56$ in code
units. The initial field strength varies between $B=0.19 - 1.77$,
corresponding to $M/M_{cr}=0.4 - 8.3$, covering the sub- and
supercritical range. The ratio $\beta=P_{th}/P_{mag}=0.01 - 4.04$.
For an overview of the parameters used see Table~\ref{tab:models}.
There are four series of models: ${\cal D}$ (pure hydrodynamics, same runs
as in Paper I); ${\cal E}$ (MHD models with the same parameter set as
the hydro models); ${\cal F}$ (magnetostatic models for numerical
tests); and ${\cal G}$ (MHD models with a reduced number of Jeans
masses). The second letter in the model names stands for the
resolution ({\em l}ow, $64^3$; {\em i}ntermediate, $128^3$; or {\em
h}igh, $256^3$), followed by a digit denoting the driving scale. For
the MHD models, a third letter gives the relative field strength ({\em
l}ow, {\em i}ntermediate, {\em m}oderate, {\em s}trong).
 
The dynamical behavior of isothermal, self-gravitating gas
depends only on the ratio between potential and kinetic
energy. Therefore we can use the same scaling prescriptions as in Paper I,
defining the physical time scale by the free-fall time 
(eq.~\ref{equ:tff}), the length scale by the initial Jeans length
\begin{equation}
\lambda_J = c_{\rm s} (\pi/(G \bar{\rho}))^{1/2} \approx (0.71 \mbox{ pc})
 \hat{\lambda}_J
 \left(\frac{c_{\rm s}}{0.2 \mbox{ km s}^{-1}}\right)
 \left(\frac{n}{10^3 \mbox{ cm}^{-3}}\right)^{-1/2},
\end{equation}
and the mass scale by the Jeans mass
\begin{equation}
M_J = \bar\rho \lambda_J^3 \approx (20.8  M_{\odot}) 
    \hat{M}_J
    \left(\frac{c_{\rm s}}{0.2 \mbox{ km s}^{-1}}\right)
    \left(\frac{n}{10^3 \mbox{ cm}^{-3}}\right)^{-1/2},
\end{equation}
where $\hat{\lambda}_J$ and $\hat{M}_J$ are expressed in code units. 
We include the magnetic field $B$ via the pressure, using 
$P=\rho c_{\rm s}^2 = B^2/(8\pi)$. This yields 
\begin{equation}
B = (4.44\times10^{-6}G)\hat{B}
           \left(\frac{c_{\rm s}}{0.2 \mbox{ km s}^{-1}}\right) 
           \left(\frac{n}{10^3 \mbox{ cm}^{-3}}\right)^{1/2},
\end{equation}
with $\hat{B}$ again in code units. 

\section{Numerical Resolution Criterion\label{sec:resolution}}

We must consider the resolution required to accurately follow
gravitational collapse.  To follow fragmentation in a grid-based,
hydrodynamical simulation of self-gravitating gas, the criterion given
by Truelove et al. (1997) holds. They studied fragmentation in
self-gravitating, collapsing regions and found that the mass contained
in one grid zone must remain significantly smaller than the local Jeans
mass throughout the computation to accurately follow the
fragmentation.  Bate \& Burkert (1997) found a similar criterion 
for particle methods.  Applying it strictly would
limit our simulations to the very first stages of collapse.
We therefore only apply this criterion to the resolution of initial
collapse.  Thereafter, we only study the gross properties of collapsed
cores, such as mass and location, but not their underresolved internal
details. 

The Truelove analysis does not include magnetic fields, which must
also be sufficiently resolved to determine whether initial collapse
occurs. Numerical diffusion can reduce the support provided
by a static or dynamic magnetic field against gravitational collapse.
Increasing the numerical resolution decreases the scale at which
numerical diffusion acts.  In this section we attempt to determine the
resolution necessary to adequately resolve magnetic support against
collapse in the presence of supersonic turbulence.  Two regimes of
field strength concern us.  For strong, subcritical fields, the
resolution should ensure that numerical diffusion remains unimportant
even for the dense, shocked regions (subsection
\ref{subsec:static-case}).  For weaker, supercritical fields, the
resolution should enable us to evolve MHD waves within the
shocked regions (subsection \ref{subsec:dynamic-case}).

\subsection{Numerical Diffusion in Magnetostatic Configurations
            \label{subsec:static-case}} 

\begin{figure}
\includegraphics{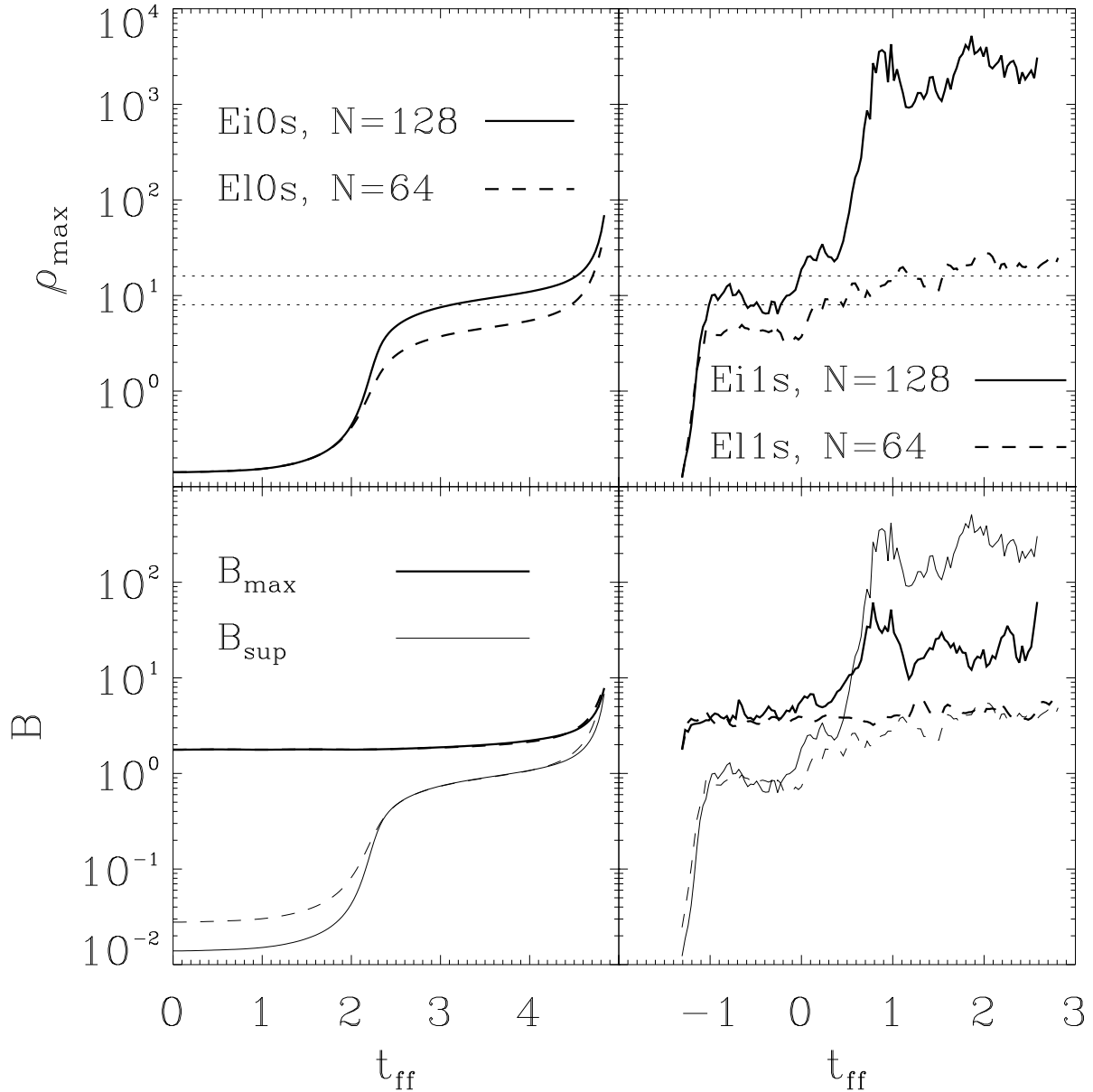}
\caption{\label{fig:numdiff}
Peak densities and maximum magnetic field strengths for strong field
($M/M_{cr}=0.8$), driven (${\cal E}l1s$ and ${\cal E}i1s$) and
undriven (${\cal E}l0s$ and ${\cal E}i0s$) runs. The dotted lines
(upper $N=128^3$, lower $N=64^3$) denote the sheet densities, i.e the
densities corresponding to all mass concentrated in a layer of one
grid zone's height. Gravity is turned on at $t=0.0$ with $t$ in units
of the global free-fall time scale $t_{\rm ff}$. The time interval
at $t < 0.0$ is necessary for the driven models to reach a state of
fully developed turbulence.  In the lower panels, the thin lines
denote the magnetic field strength required to support a region of
density $\rho_{max}$ according to equation \ref{equ:magstatsup}}
\end{figure}

We begin by considering magnetostatic
support. Figure~\ref{fig:numdiff} demonstrates that numerical
diffusion can dominate the behavior in this case.  The left panel
displays the peak density $\rho_{max}$ and magnetic field amplitude
$B_{max}$ for the low and intermediate resolution undriven models
${\cal E}l0s$ and ${\cal E}i0s$, while the right panel contains the
same quantities for the driven, but otherwise identical, models ${\cal
E}l1s$ and ${\cal E}i1s$ (see Table~\ref{tab:lambda-j}).  All these
models have initial magnetic field strength sufficient to support the region,
with $M/M_{cr}=0.8$.

Starting with a sinusoidal density perturbation in the undriven case, both 
the driven and undriven models first collapse into sheet structures.  
The dotted lines in the density plots show the density $\rho_{\rm sheet}$
corresponding to having all the mass in the box in a layer one zone
thick. In a volume filled with otherwise unperturbed gas of initially
uniform density, reaching peak densities $\rho_{max}$ higher than this
threshold means that numerical diffusion across field lines must have
begun, as can be seen in the top left panel of
Figure~\ref{fig:numdiff}.  However, in driven, isothermal, supersonic
turbulence, the peak densities in shocked regions scale with the Mach
number ${\cal M}$ as $\rho_{pk}\propto {\cal M}^2$. Thus, $\rho_{pk}$
can reach values orders of magnitude higher than the mean density,
easily exceeding $\rho_{sheet}$ even in the absence of diffusion.

The magnetic field provides an alternate diagnostic. In the
absence of numerical diffusion, mass should be tied to the field lines
running vertically through the cube, so the mass to flux ratio along
any given field line should not change.  In the lower panels, we plot
the magnetic field strength $B_{sup}(\rho)$ required to support a
region of density $\rho_{max}$ (thin lines) according to equation
\ref{equ:magstatsup}.  The magnetic field starts out significantly
stronger than $B_{sup}$, as the mass is insufficient for collapse to
occur.  If $B_{max}$ grows more slowly than $B_{sup}$, then density
must be diffusing across field lines.  If the two values cross, then
numerical diffusion has allowed collapse to occur unphysically.  This
happens in the undriven models at $t=4.7t_{\rm ff}$, while in the
driven models with their greater density contrasts, collapse already 
occurs at $t=0.5t_{\rm ff}$.  We conclude that in these low and 
intermediate resolution models, the resolution is not
sufficiently high for the magnetic field to follow the turbulence,
especially in shocked regions. 

\begin{figure}
\epsfig{file=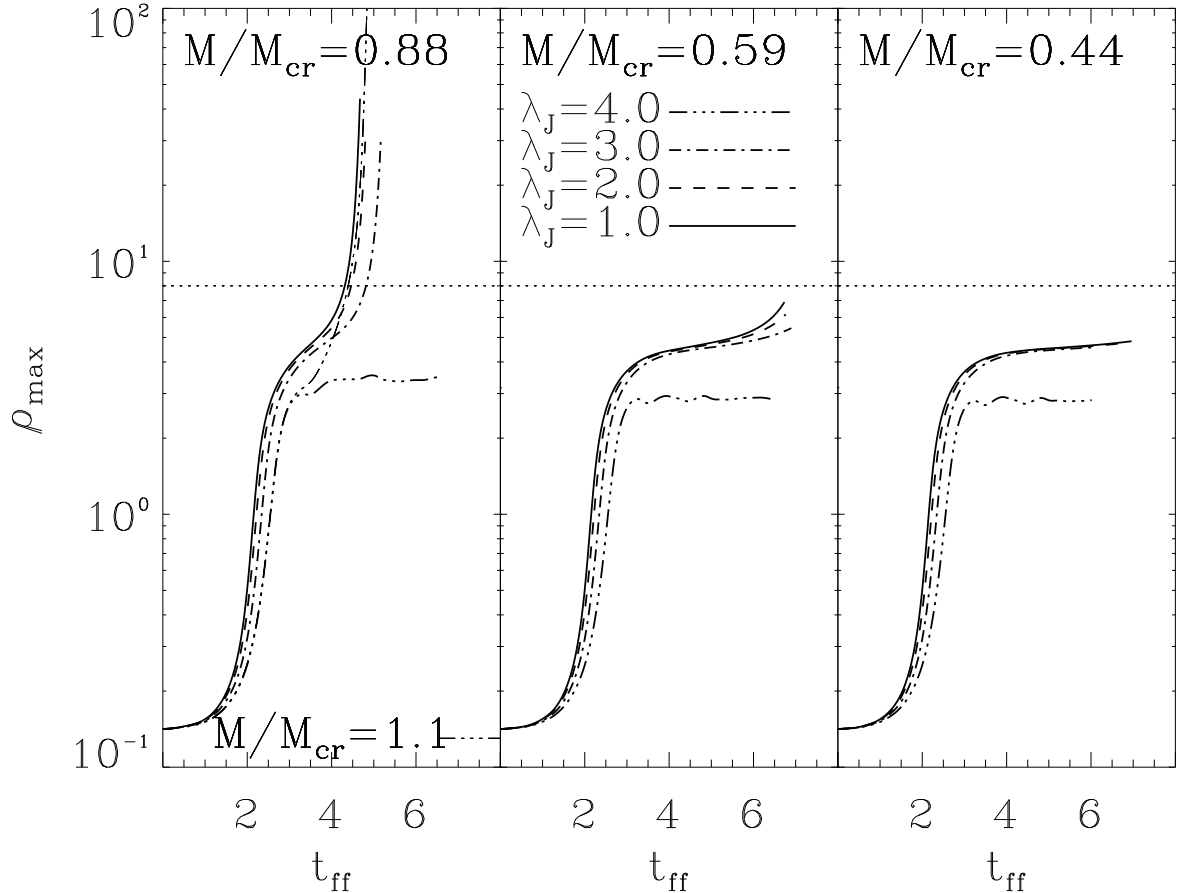,width=16.0cm}
\caption{\label{fig:testsuite-magstat}
Peak densities for all test models of type ${\cal F}$. 
 $\lambda_J$ is the Jeans length in units
of grid zones, when all the mass is collected in a sheet of a single zone's
height. The horizontal dotted line denotes the density reached if all
mass is collected in sheet of a single zone's height.
Note that for $\lambda_J=4.0$ zones we get collapse in the 
supercritical case ($M/M_{cr}=1.1$), but not in the subcritical one. Thus,
we verify the mass-to-flux criterion (equation \ref{equ:magstatsup}).}  
\end{figure}

We can use this example to derive a criterion for the resolution of a
magnetostatically supported sheet.  We can also confirm that we are
using the correct numerical constant in the mass-to-flux criterion
given by equation (\ref{equ:magstatsup}).  We did a suite of models
${\cal F}$ (see Table \ref{tab:models}) varying the sound speed and
the mass in the cube while holding the magnetic field strength
constant, thus varying $M/M_{cr}$ and the number of zones in a Jeans
length $\lambda_J$.  For $\lambda_J=1.0$, we certainly cannot expect
to get reliable results, as this Jeans length does not even fulfill
the hydrodynamic criterion of Truelove et al. (1997).  In
Figure~\ref{fig:testsuite-magstat} we present the results.  We find
unphysical collapse occurring for physically supported regions until
$\lambda_J \geq 4.0$ zones.  We also find that at this resolution, a
model with $M/M_{cr} = 1.1$ collapses (thin line in first panel) while
a model with $M/M_{cr} = 0.88$ does not collapse, confirming equation
\ref{equ:magstatsup}.  We conclude that {\em for a self-gravitating
magnetostatic sheet to be well resolved, its Jeans length must exceed
four zones}.

\subsection{MHD Waves in High Density Regions 
            \label{subsec:dynamic-case}}

As we want to investigate whether MHD turbulence can prevent
gravitationally unstable regions from collapsing, we have to be sure
to resolve the MHD waves, which are the main agent in this mechanism.
The same argument holds as for the criterion to prevent local collapse
in the pure hydro case (Paper I): If the wavelength of the turbulent
perturbation is smaller than the {\em local} Jeans length $\lambda_J$,
stabilization should be possible, at least in principle. To sample a
sinusoidal wave on a grid, we need at least four zones.
Polygonal interpolation of a sine wave with evenly spaced supports
would then yield an error of $\approx 21$\%.  This decreases to
$\approx 9$\% with six zones and $\approx 5$\% using eight zones.  We
choose to set the minimum permitted Jeans length to six zones,
admittedly somewhat arbitrarily.

In table \ref{tab:lambda-j} we list the local Jeans lengths for all
model types and their resolution. Models of type ${\cal E}$ start to
be resolved at a resolution of $N=256^3$, whereas the ones of type
${\cal G}$ can already be regarded as resolved at $N=128^3$.

\section{Results \label{sec:results}}

In the previous section we considered under what circumstances
numerical effects could allow unphysical gravitational collapse.  In
this section we now consider adequately resolved models in order to
determine whether magnetized turbulence can prevent the collapse of
regions that are not magnetostatically supported.  We begin by
demonstrating that supersonic turbulence does not cause a
magnetostatically supported region to collapse, and then demonstrate
that in the absence of magnetostatic support, MHD waves cannot
completely prevent collapse, although they can retard it.

\subsection{Magnetostatic Support\label{subsec:subcrit}}

\begin{figure}
\epsfig{file=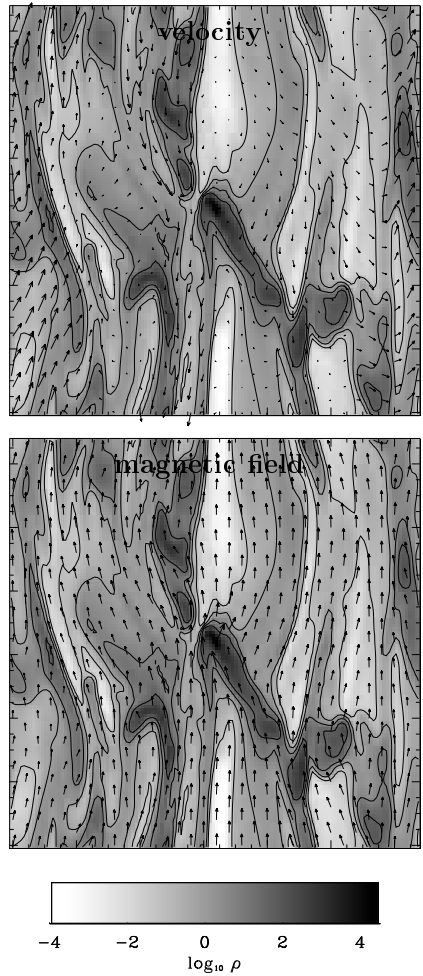,height=18.0cm}
\caption{\label{fig:magstatsup}
Two dimensional slice of increased Jeans mass model ${\cal G}i1s$ with velocity 
field vectors (upper panel) and magnetic field vectors (lower panel), displaying
the wohle computational domain. 
The initial magnetic field is oriented along the $z$-direction,
i.e. vertically in all plots presented. Driving happens at $k=1-2$.
The field is strong enough in this case not only to prevent 
the cloud from collapsing perpendicular to the field lines, 
but even suppress the turbulent motions in the cloud. The 
turbulence only scarcely affects the mean field. The picture
is taken at $t=5.5t_{\rm ff}$.}
\end{figure}

In a subcritical region with $M < M_{cr}$, the cloud is expected to
collapse to a sheet, which in turn should be stable. Figure
\ref{fig:magstatsup} shows the corresponding model ${\cal G}i1s$.
These runs have been computed with a lower Mach number ${\cal M} =
5.0$ in order to demonstrate the behavior of a magnetostatically
supported cloud.  The initially uniform magnetic field runs parallel
to the $z$-axis.  The field is strong enough to force significant
anisotropy in the flow, although the dense sheets that form do not
always lie perpendicular to the field lines as the driving can shift
the sheets along the field lines without changing the mass-to-flux 
ratio. The sheets do not collapse further, because the shock waves 
cannot sweep gas across field lines and the cloud is initially 
supported magnetostatically.

\begin{figure}
\includegraphics{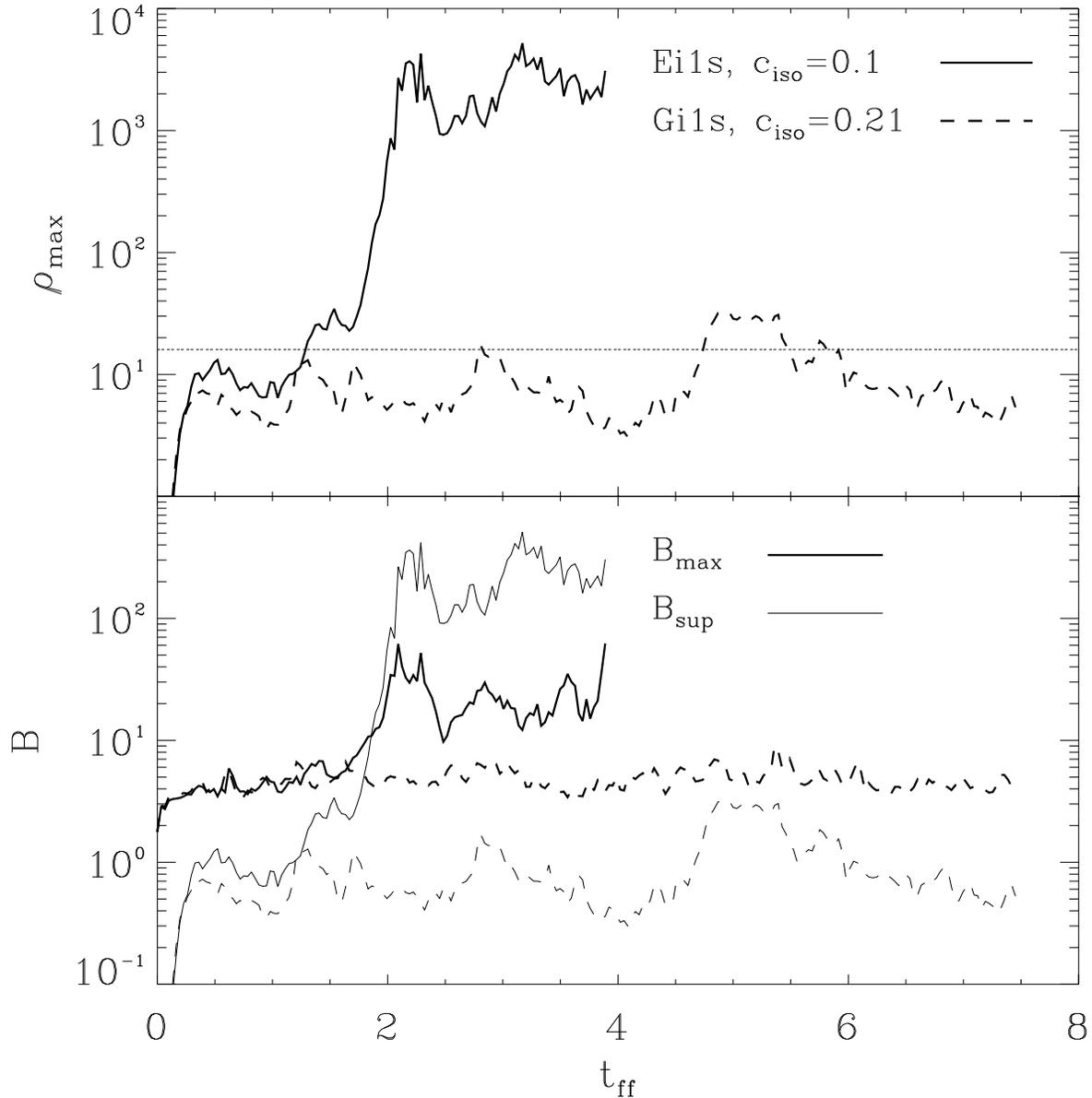}
\caption{\label{fig:numdiffcheck}
Peak densities and maximum magnetic field strengths for strong field
model ${\cal E}i1s$ (solid line) and model ${\cal G}i1s$ (dashed line). 
The dotted line denote the sheet density, i.e the density corresponding 
to all mass concentrated in a layer of one grid zone's height. Gravity is 
turned on at $t=0.0$ with $t$ in units of the global free-fall time. 
In the lower panel, the thin lines denote
the magnetic field strength required to support a region of 
density $\rho_{max}$ according to equation \ref{equ:magstatsup}.
Whereas model ${\cal E}i1s$ shows unphysical, numerical collapse,
model ${\cal G}i1s$ is well resolved.}
\end{figure}

Figure \ref{fig:numdiffcheck} demonstrates that this result is
reasonably well resolved numerically.  As in figure \ref{fig:numdiff},
we show peak densities and magnetic field strength for two models that
differ only in their sound speeds, and thus by the number of zones in
a Jeans wavelength $\lambda_J$.  Whereas model ${\cal E}i1s$, with
$\lambda_J = 3.2$ zones does collapse, although it should be
supported, model ${\cal G}i1s$, with $\lambda_J= 6.8$ zones behaves as
expected physically, just as our resolution criterion predicts.
Note that the actual field strength in model
${\cal G}i1s$ always exceeds $B_{sup}$, the field strength necessary
to support the region.

\subsection{MHD Wave Support\label{subsec:supercrit}}

A supercritical cloud with $M > M_{cr}$ is not magnetostatically
supported, and could only be stabilized by MHD wave pressure, assuming
ideal MHD.  In this section we show that this appears to be
insufficient to completely prevent gravitational collapse, although it
can slow the process down.

\subsubsection{Morphology \label{subsubsec:morph}}

\begin{figure}
\epsfig{file=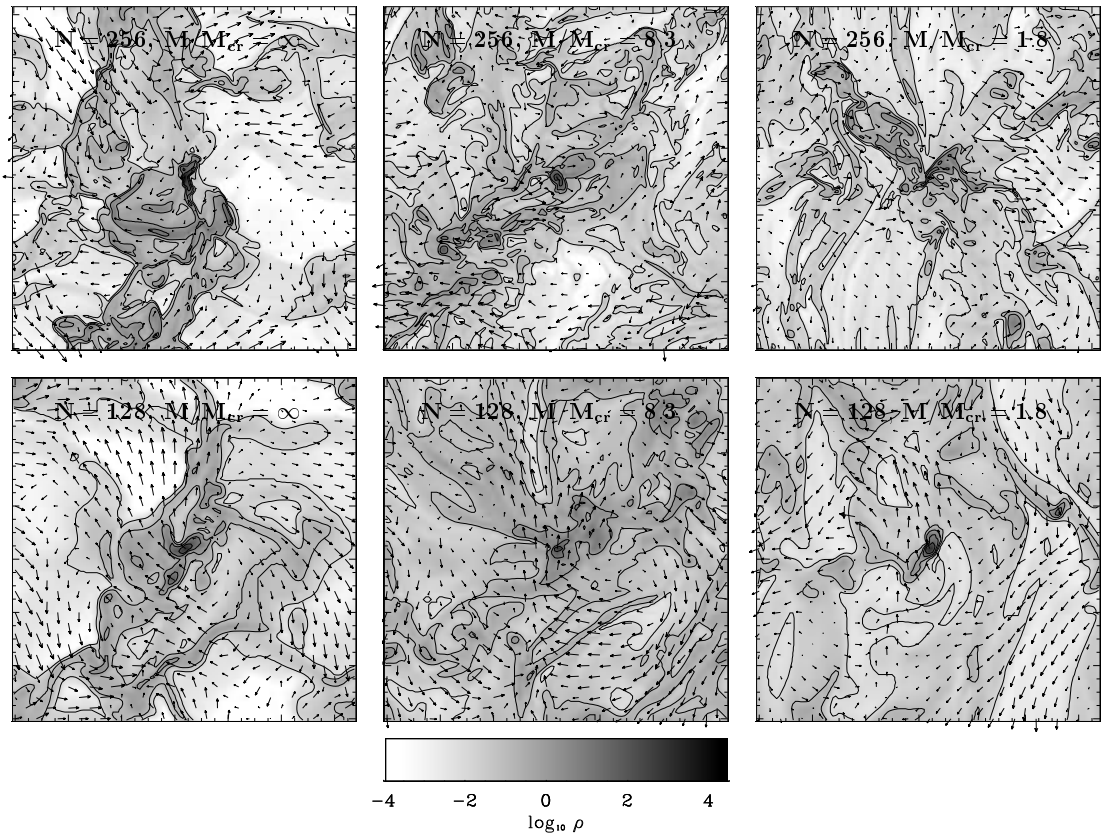,width=16.0cm}
\caption{\label{fig:2Dslices}
Two-dimensional slices of the high-resolution models ${\cal D}h1$,
${\cal E}h1w$ and ${\cal E}h1i$ and the corresponding models at
intermediate resolution ${\cal D}i1$, ${\cal E}i1w$, and ${\cal E}i1i$,
displaying the whole computational domain. 
Slices are taken at the location of the zone with the highest
density at the time when $10$\% of the total mass has been accreted
onto cores. The plot is centered on this zone.  Arrows denote
velocities in the plane. The length of the largest arrows corresponds
to a velocity of $v \sim 2.0$. Greyscale stands for density, where highest
density regions are darkest. All slices are scaled equally, using the
same scale as \ref{fig:magstatsup}.
Driving, as in figure \ref{fig:magstatsup}, happens at $k=1-2$.}
\end{figure}

In the upper panels of Figure \ref{fig:2Dslices} we compare the
morphology of hydrodynamical (${\cal D}h1$), weakly magnetized (${\cal
E}h1w$), and strongly magnetized (${\cal E}h1i$) supercritical models
at a resolution of $256^3$ zones.  The figure presents two dimensional
slices through the three dimensional simulation volume centered on the
locations of the most massive clumps. In order to compare the models
at similar stages of their evolution, we took the snapshots at a time
when roughly 10\% of the total mass had been accreted onto collapsing clumps.
All three runs show well developed turbulence, rarefied regions,
shocked regions, and at least one clump. However, model ${\cal E}h1w$,
with $M/M_{cr} = 8.3$, seems to contain more power on small scales
than the pure hydro run, model ${\cal D}h1$ ($M/M_{cr} = \infty$).  We
discuss this more quantitatively below. In model ${\cal E}h1i$, with
$M/M_{cr}=1.8$, the vertically oriented mean field (in the plane of
the Figure) starts producing some anisotropy.  This model represents a
morphological transition from the pure hydrodynamical model ${\cal
D}h1$ with completely randomly oriented motions to the
magnetostatically supported model ${\cal G}i1s$, with its ordered
strucures.

\subsubsection{Resolution Study\label{subsubsec:resolution}}

We must address the question of whether our magnetodynamic simulations
are indeed well resolved.  The parameters for models ${\cal E}$ yield a
global Jeans length of $\lambda_J \approx 0.5$, corresponding to a
local, post-shock Jeans length of $\lambda_J \approx 0.05$ for
isothermal shocks with Mach number ${\cal M} \approx 10$ (see
Table~\ref{tab:lambda-j}).  At $N=128^3$ zones, this results in a
local Jeans length of only $3.2$ zones, but at $256^3$ the local Jeans
length is $6.4$ zones, satisfying our resolution criterion.  Instead
of increasing the resolution, we increased the Jeans length in the
models ${\cal G}$ discussed below in \S~\ref{subsubsec:ratioes} by
increasing the sound speed. In these models we used a global Jeans
length of $\lambda_J \approx 1.1$, corresponding to a local Jeans
length of $6.8$ zones at $N=128^3$.

\begin{figure}
\includegraphics{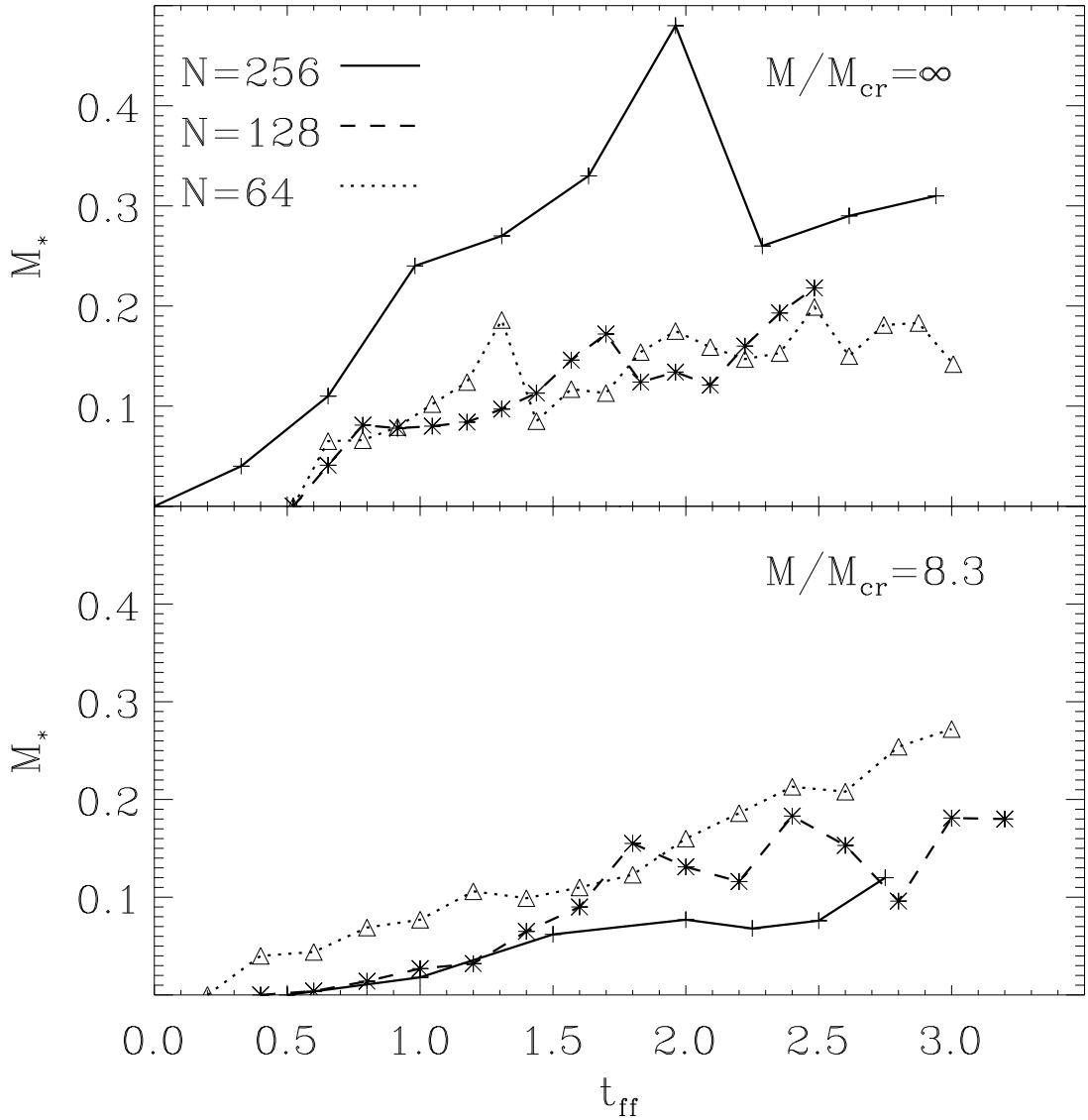}
\caption{\label{fig:massfrac-res}
Comparison of the mass accretion behaviour for runs driven at $k=1-2$
with varying resolution. Pure hydro runs are shown in the upper panel
(models ${\cal D}l1$ (dotted), ${\cal D}i1$ (dashed), and ${\cal D}h1$
(solid)), and MHD runs are shown in the lower panel (models ${\cal
E}l1$ (dotted), ${\cal E}i1$ (dashed), and ${\cal E}h1$ (solid)).
$M_*$ denotes the sum of masses found in all cores determined by
the modified CLUMPFIND (Williams et al. 1994, see also Paper I). Times 
are given in units of free-fall time as defined in equation \ref{equ:tff}. 
Although the collapse rate varies, we get collapse in all cases.}
\end{figure}

Figure~\ref{fig:2Dslices} compares high resolution $256^3$ zone models
with the corresponding lower resolution $128^3$ zone models in the
same dynamical state, when 10\% of the mass has been accreted onto
cores.  The lower resolution makes itself felt in broader shocks in
all cases, so that the peak densities are lower than in the
high-resolution runs.  In the MHD models decreasing the resolution
also leads to thicker collapsed sheets.  Thus, unstable regions form
at later times in both the MHD and hydrodynamical cases, as can be
seen in Figure \ref{fig:massfrac-res}.  There, we show the core mass
accretion history for three weakly magnetized models varying only in
their resolution ${\cal E}l1w$ ($64^3$), ${\cal E}i1w$ ($128^3$) and
${\cal E}h1w$ ($256^3$) (lower panel) and the corresponding
hydrodynamical models (${\cal D}l1$, ${\cal D}i1$, ${\cal D}h1$)
(upper panel).  Cores were determined using the modified CLUMPFIND
algorithm of Williams et al. (1994) as described in Paper I.

Collapse occurs in both cases at all resolutions. However, increasing
the resolution makes itself felt in different ways in hydrodynamical
and MHD models.  In the hydrodynamical case, higher resolution results
in thinner shocks and thus higher peak densities.  These higher
density peaks form cores with deeper potential wells that accrete more
mass and are more stable against disruption.  If we increase the
resolution in the MHD models, on the other hand, we can better follow
short wavelength MHD waves, which appear to be able to delay collapse,
although not to prevent it.

\begin{figure}
\includegraphics{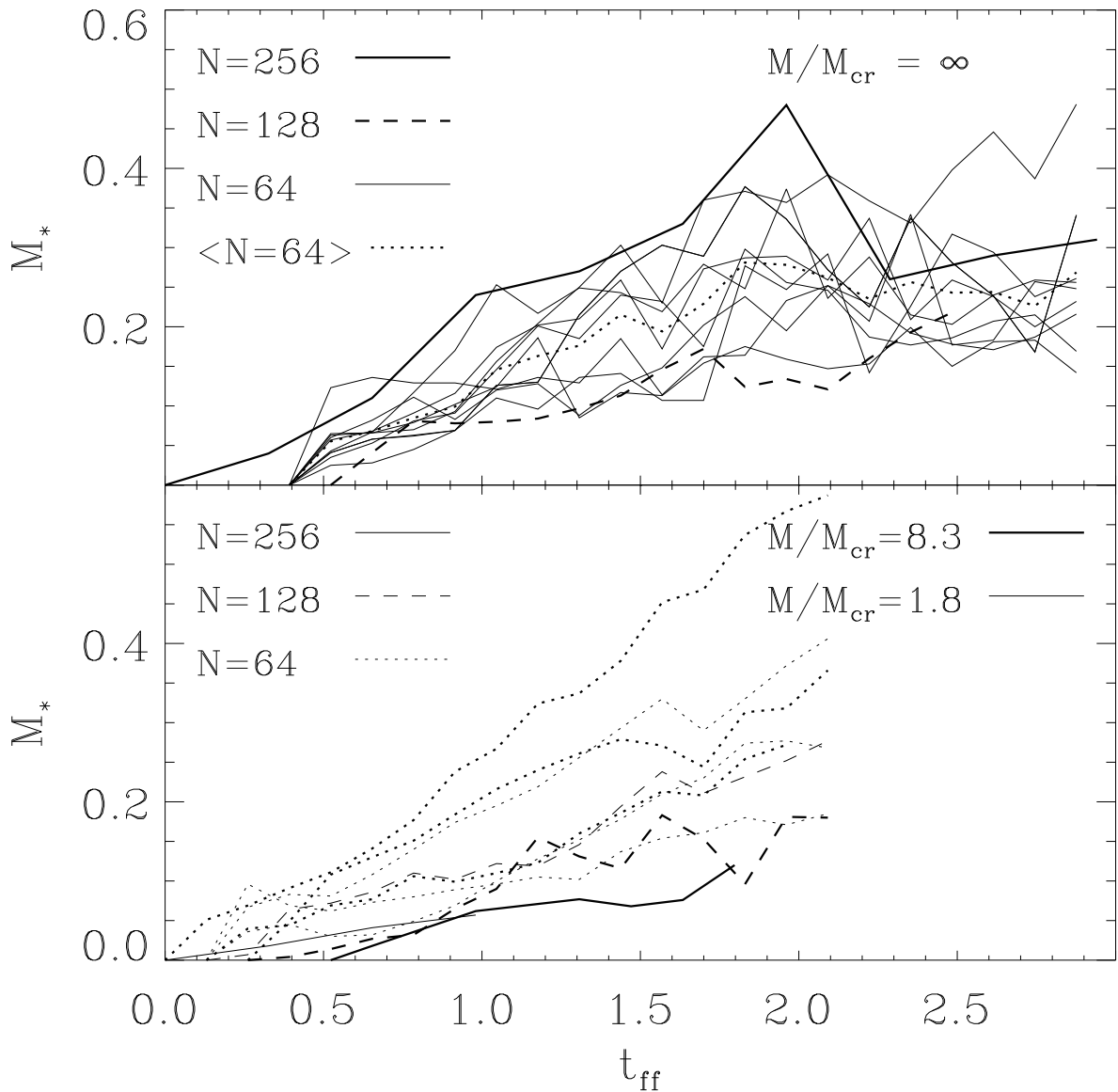}
\caption{\label{fig:variance}
(Upper panel) Core-mass accretion rates for $10$ hydro runs with equal
parameter set (model ${\cal D}l1$) but different realizations of the
turbulent velocity field. The thick line shows a ``mean accretion
rate'', calculated from averaging over the sample. For comparison, the
higher-resolution runs ${\cal D}i1$ and ${\cal D}h1$ are shown. The
latter one ($N=256^3$) can be regarded as an envelope for the low
resolution models.  (Lower panel) Mass accretion rates for runs with
identical magnetic fields but different driving fields, and runs with
identical driving fields but different magnetic fields (models 
${\cal E}l1w$, ${\cal E}i1w$, ${\cal E}h1w$ and ${\cal E}l1i$, 
${\cal E}i1i$, ${\cal E}h1i$).  The effects of magnetic fields are
covered by variations due to the turbulent velocity field. Identical
line styles stand for models with identical parameters, but different
driving velocity fields. Models ${\cal E}l1w$ and ${\cal E}l1i$ have 
been computed $3$ times with varying driving velocity fields.} 
\end{figure}

The turbulent formation of gravitationally condensed regions via shock
interactions is a highly stochastic process.  As in Paper I, we
demonstrate this by choosing different random realizations of the
driving velocity field with the same characteristic
wavelengths. Figure~\ref{fig:variance} shows the core mass accretion history
for a pure hydrodynamical model set and a MHD set. In the upper panel, we
plotted the models ${\cal D}l1$, ${\cal D}i1$, ${\cal D}h1$, where the
low-resolution model ${\cal D}l1$ has been repeated multiple times (solid thin
lines).  The dotted thick line denotes the average of these
runs.  We find that resolution effects are exceeded by statistical
variations caused by random variations of the driving fields.

In the MHD case (lower panel) the thickness of the lines stands for
the strength of the field, expressed in terms of the ratio
$M/M_{cr}$.  Dotted lines denote low-resolution runs computed
with varying driving velocity fields, as in the upper panel.  The
high-resolution run with $M/M_{cr}=1.8$ was stopped at $t=1.0t_{\rm ff}$
because the Alfv\'{e}n timestep became prohibitively small.  Increasing
the resolution makes itself felt for the runs with stronger fields
in the same way as for the ones with weak fields: The higher the
resolution, the better the small-scale MHD waves are resolved, and
thus the slower the collapse.  Collapse does always occur, however.

\subsubsection{Core Distribution\label{subsubsec:structure}}

\begin{figure}
\epsfig{file=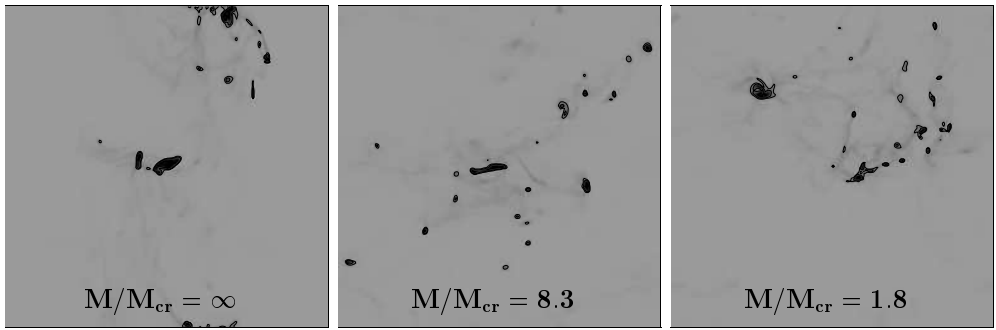,width=15.0cm}
\caption{\label{fig:corecoords}
Projected coordinates of clumps found when $10$\% of the total mass has been
accreted onto cores. All simulations (models ${\cal D}h1$, ${\cal E}h1w$ and
${\cal E}h1i$) are driven at wavenumbers $k=1-2$. 
For the pure hydro case, we get strongly clustered collapse, whereas for
supercritical fields, the cores are more evenly distributed. For 
slightly supercritical fields ($M/M_{cr}=1.8$, model ${\cal E}h1i$), the 
cloud tends to collapse along the field lines, so that the extent of
the core distribution is reduced in direction parallel to the initial
field (vertically oriented in the plots).} 
\end{figure}

Although MHD waves cannot prevent local collapse entirely, the
resulting collapse 
appears qualitatively different from collapse in the
hydrodynamic case with corresponding global $\lambda_J$ and driving
strength.  
In the hydrodynamical case driven at  $\lambda_D > \lambda_J$, shocks
are widely separated and sweep up substantial mass, producing isolated
clusters of cores.
In the presence of a weak
supercritical field, 
the shock structure appears to have more small-scale structure,
resulting in cores distributed more
uniformly across the simulation volume, as shown in the middle panel
of Figure~\ref{fig:corecoords}.  In fact, the weakly magnetized model
driven on large scales with $\lambda_D > \lambda_J$ rather more
resembles the hydro model driven on small scales with $\lambda_D <
\lambda_J$ shown in Figure~11 of Paper I.  

\begin{figure}
\includegraphics{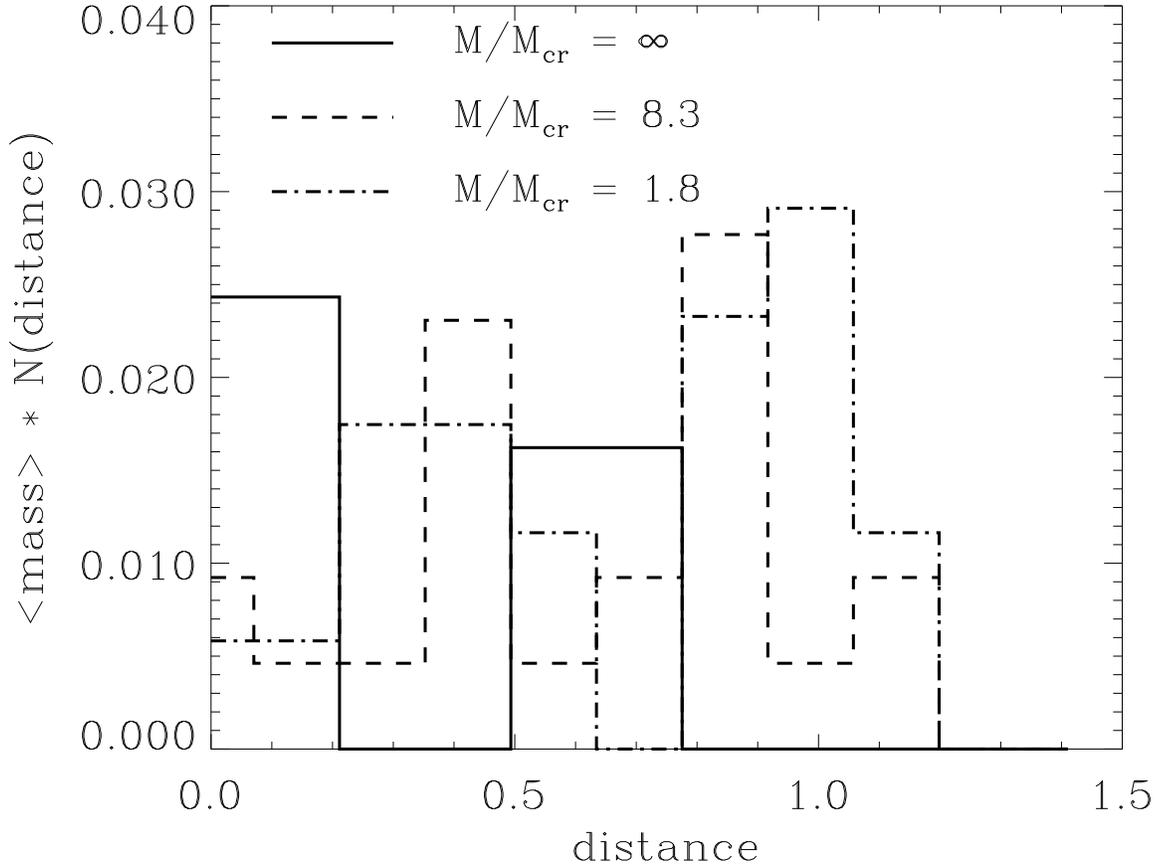}
\caption{\label{fig:r-dist}
Histogram of core distances weighted with the mean core mass for the
projected cores of Figure~\ref{fig:corecoords}. The box length is
$L=2$. The weighted means and their standard deviations are
shown in Figure~\ref{fig:b-r-dist}. Although the statistics are
not sufficient, the magnetized models tend to show a more uniform
distribution.
}
\end{figure}

\begin{figure}
\includegraphics{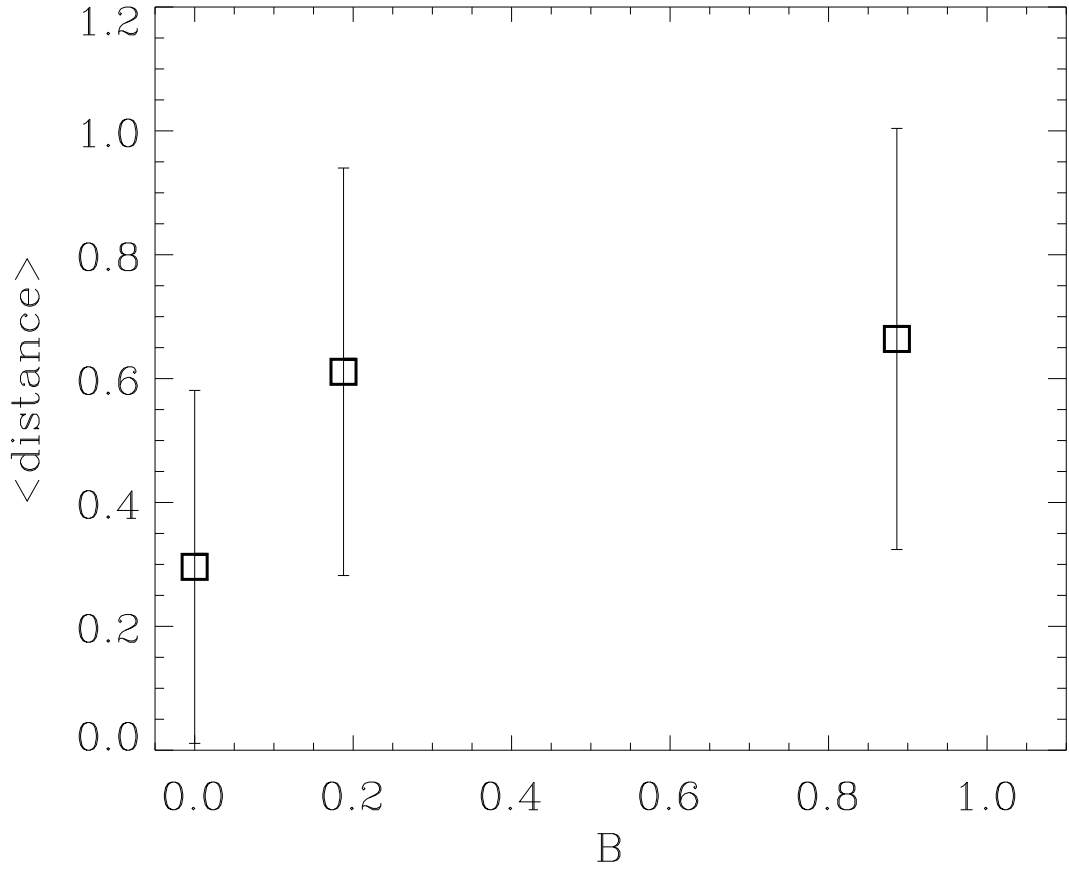}
\caption{\label{fig:b-r-dist}
Weighted means and their standard deviations for the core distances of
Figure~\ref{fig:r-dist}. The effect of low number statistics is
clearly to be seen (between six and ten cores were found).
}
\end{figure}

Figures \ref{fig:r-dist} and \ref{fig:b-r-dist} try to
quantify this difference. Figure \ref{fig:r-dist} shows the 
histogram of core distances for each panel of Figure~\ref{fig:corecoords},
multiplied with the mean core mass. A clustered ensemble of 
high-mass cores should result in a peaked distribution at small distances,
whereas a spread-out ensemble of low-mass cores should have a broader 
distribution. Figure~\ref{fig:r-dist} hints at such a behaviour,
although we are well aware of the fact, that the statistics barely 
suffice. Nevertheless, the distribution for the magnetized runs is 
shifted to larger radii. Note that the total mass in the cores found is 
within $10$\% the same for all three models. 
Figure~\ref{fig:b-r-dist} shows the weighted means of core distances, with 
their standard deviations as error bars.  Again, we see a slight shift
towards larger separations, suggesting a more uniform distribution of
cores in the MHD cases, although larger simulations with greater core
numbers will be needed to confirm this result.

\subsubsection{Energy Distribution\label{subsubsec:ratioes}}

\begin{figure}
\epsfig{file=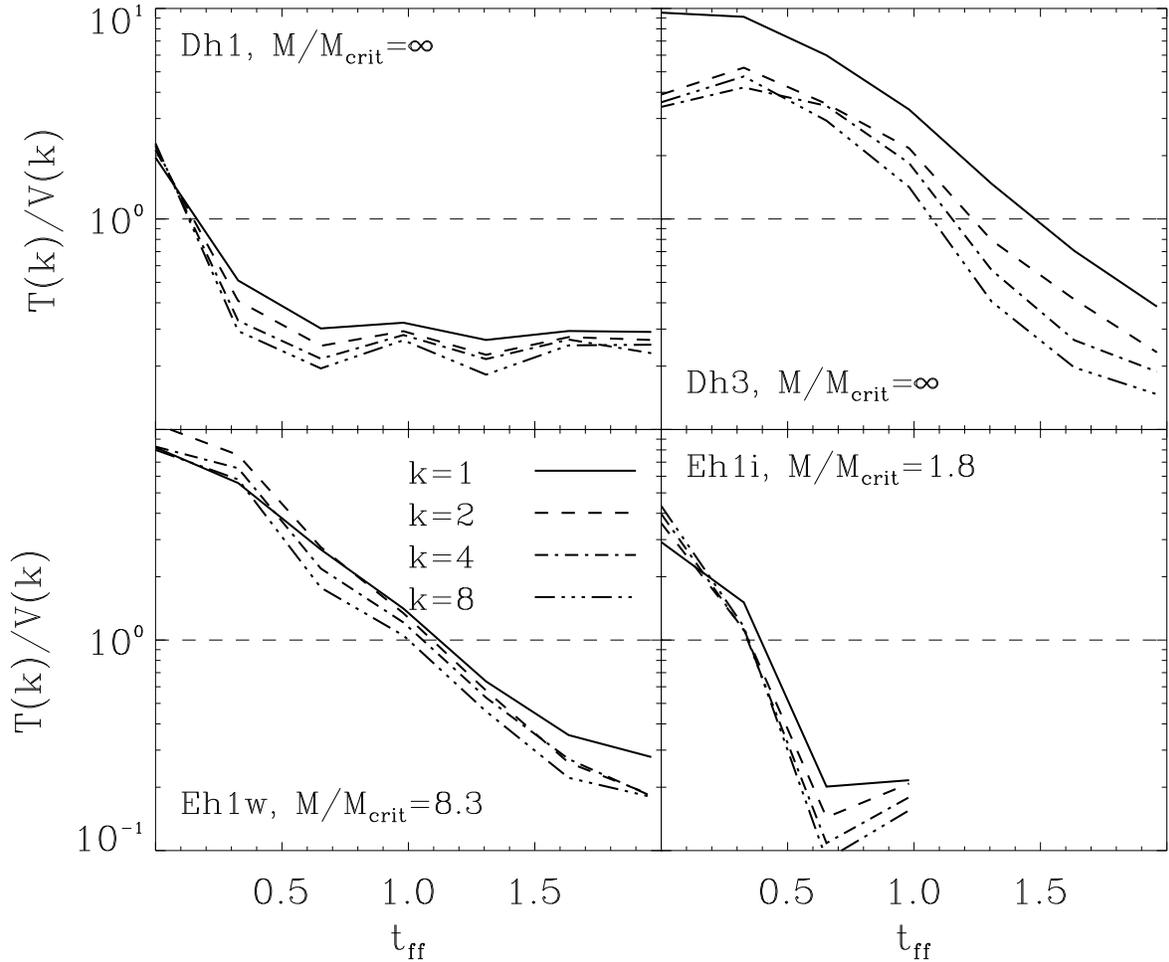,height=16.0cm,angle=90}
\caption{\label{fig:kmodes}
Time evolution of the ratio of kinetic to potential energy, $T/V$
split up according to wavenumber $k=1,2,4,8$ (solid, dashed,
dash-dotted and dotted lines) for models ${\cal D}h1$, ${\cal D}h3$
(driven at $k=7-8$), ${\cal E}h1w$ and ${\cal E}h1i$. The horizontal
line at $T/V=1.0$ indicates the instability boundary. Time is normalized
to units of global free-fall time $t_{ff}$. Note that a strong subcritical
field leads to faster collapse than a weak subcritical one.}
\end{figure}

Further evidence for a qualitative difference between hydrodynamic and
MHD collapse comes from Figure~\ref{fig:kmodes}.  Here, we show the
time evolution of the ratio of kinetic to potential energy, decomposed
into contributions from four spatial scales.  (The time resolution is
somewhat coarse as these models with $256^3$ zones were only dumped
every $t=0.3 t_{\rm ff}$.)  Whereas the hydrodynamic model (${\cal
D}h1$) driven at $k=1-2$ collapses within less than $0.5t_{\rm ff}$,
the weakly magnetized model ${\cal E}h1w$ ($M/M_{crit}=8.3$) is
supported until $t \gtrsim 1 t_{\rm ff}$ before also collapsing. This
is at least qualitatively similar to the behavior of model ${\cal
D}h3$, which is driven at wavenumbers $k=7-8$ and thus 
has a denser
network of shocks.  On the other hand, Figure~\ref{fig:kmodes}
suggests that the stronger-field model ${\cal E}h1i$ ($M/M_{cr}=1.8$)
collapses even more thoroughly than its hydrodynamical counterpart due
to the ordering influence of the strong mean field.

\begin{figure}
\epsfig{file=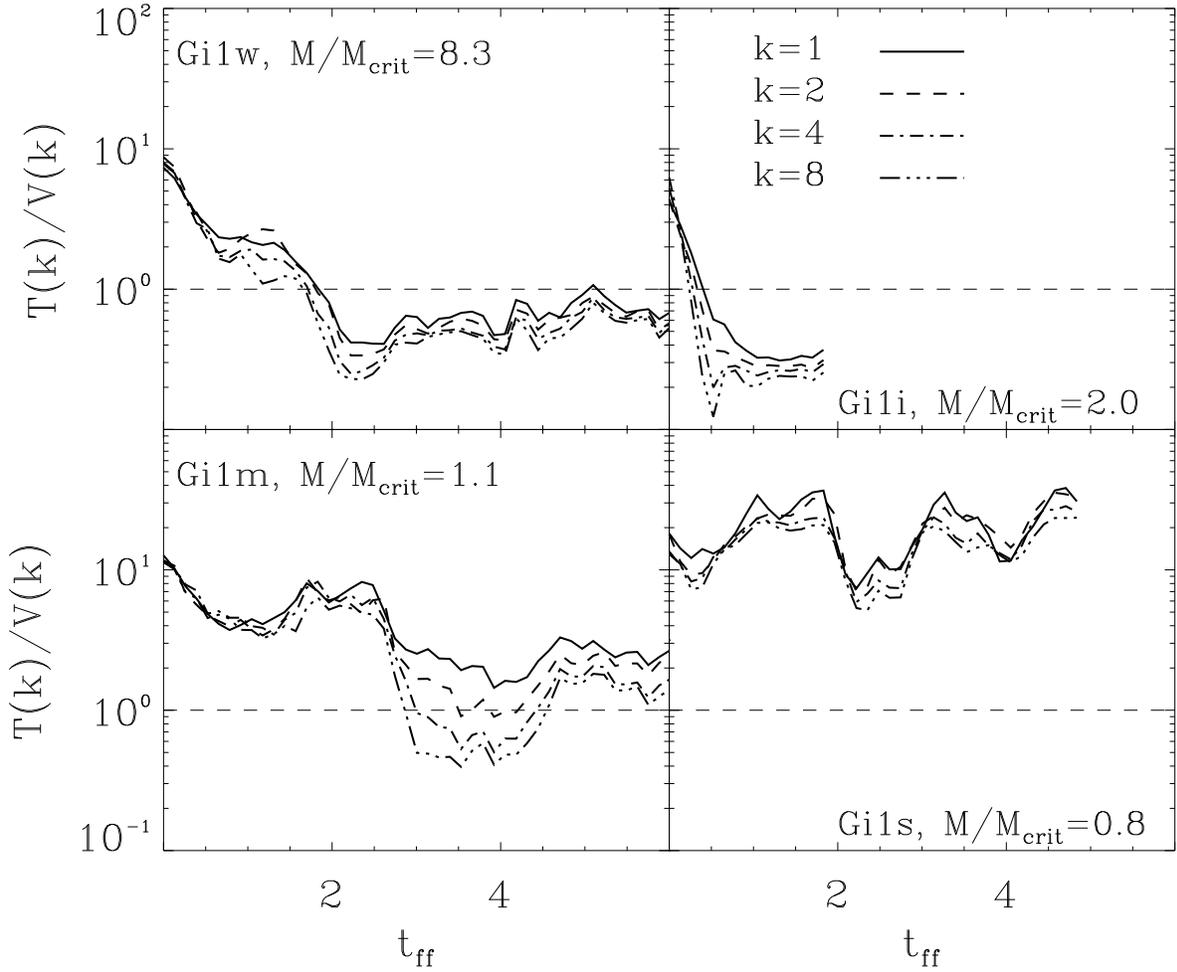,height=16.0cm,angle=90}
\caption{\label{fig:engspec-nj12}
Ratio of kinetic to potential energy against time for models ${\cal
G}$ with reduced number of Jeans masses ($n_J = 6.4$), split
up into contributions from four spatial scales ($k=1,2,4,8$).  All
models show collapse except for the magnetostatically supported one
(${\cal G}i1s$). Note that models ${\cal G}i1w$ and ${\cal G}i1i$
behave as their counterparts ${\cal E}h1w$ and ${\cal E}h1i$ shown in
figure \ref{fig:kmodes}.  }
\end{figure}

We use the model series ${\cal G}$ to follow the collapse to later
times and with more frequent time sampling. These $128^3$ models still
resolve the Jeans length in the densest clumps, as discussed in
\S~\ref{subsubsec:resolution}. We reduced the
Mach number to ${\cal M}=5$ in order to maintain an
energy input comparable to models ${\cal E}$.  This decreases the rms
post-shock density, so we actually expect the ${\cal G}$ series to
form cores with somewhat more difficulty than the ${\cal E}$ series.
Figure \ref{fig:engspec-nj12} shows the ratio of kinetic to potential
energy of the magnetized models in the ${\cal G}$ series.  (The
hydrodynamical model ${\cal G}i1$ collapses within half a free-fall
time after gravity is turned on at $t=0.0$.)  With increasing field
strength (models ${\cal G}i1w$ and ${\cal G}i1i$), the collapse is
delayed, but never prevented. This even applies for ${\cal G}i1m$,
where the field strength is only marginally supercritical
($M/M_{crit}=1.1$).  In this model, bound cores form, but are then
destroyed by passing shocks, probably for the unphysical reason that
they cannot continue collapsing to sizes smaller than a few zones (see
discussion in Paper I).  Increasing the field strength further leads
to model ${\cal G}i1s$ ($M/M_{crit}=0.8$), where the field supports the cloud
magnetostatically, and gravitationally bound cores do not form.

\subsubsection{Energy Spectra\label{subsubsec:spectra}}

\begin{figure}
\epsfig{file=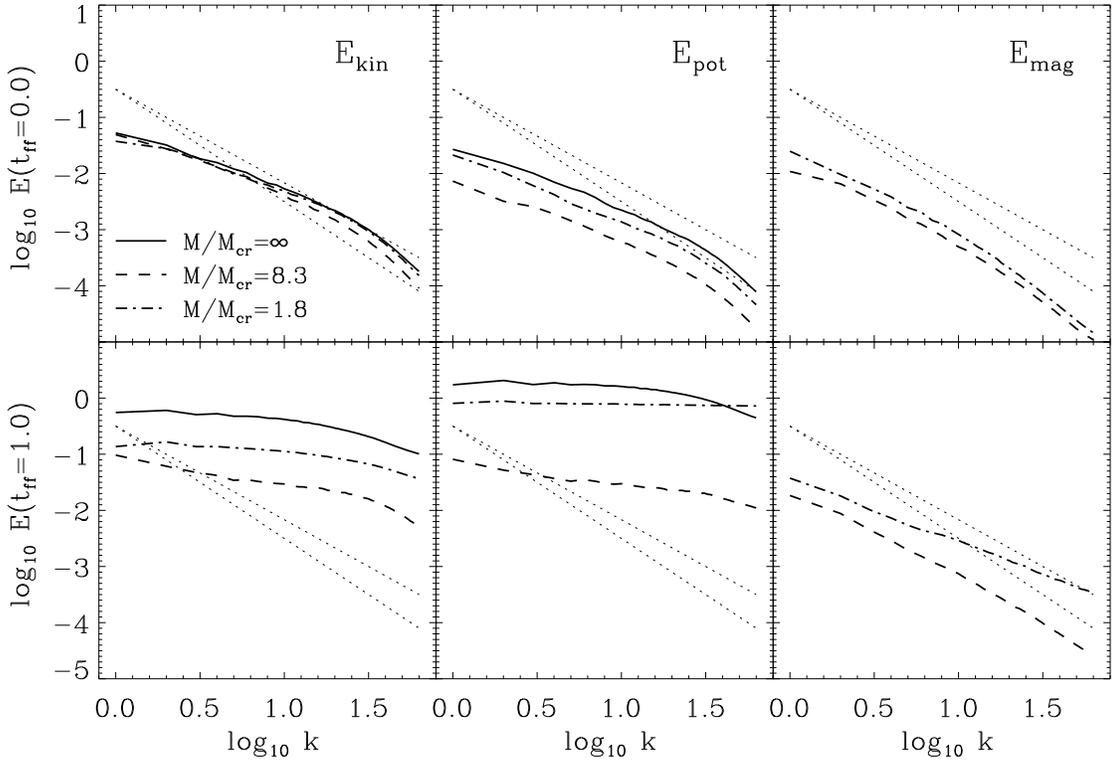,height=16.0cm,angle=90}
\caption{\label{fig:engspec-t=0}
Kinetic, potential and magnetic energies for models 
${\cal D}h1$ ($k=1-2$, $M/M_{cr}=\infty$),
${\cal E}h1w$ ($k=1-2$, $M/M_{cr}=8.3$) and ${\cal E}h1i$ ($k=1-2$,
$M/M_{cr}=1.8$) at the time $t=0.0$ at which gravity is
turned on in a state of fully developed turbulence (upper row). For
comparison we included power spectra of $P(k)\propto k^{-5/3}$ and
$P(k)\propto k^{-2}$ (dotted lines). The lower row contains the same
models, but for time $t=1.0 t_{\rm ff}$.}
\end{figure}

We next examine Fourier spectra of the energy.  Figure
\ref{fig:engspec-t=0} presents the spectra of kinetic, potential 
and magnetic energies at times $t=0.0$ and
$t=1.5t_{\rm ff}$ of the high-resolution models ${\cal D}h1$
(hydrodynamic), ${\cal E}h1w$ ($M/M_{cr}=8.3$) and ${\cal E}h1i$
($M/M_{cr}=1.8$), all driven at wavenumber $k=1-2$. The spectra at
time $t=0.0$ represent fully developed turbulence just before gravity
is switched on.  Here we find another reason for the fast collapse of
the more strongly magnetized model ${\cal E}h1i$.  The density
enhancements due to shock interactions are larger for models ${\cal
D}h1$ and ${\cal E}h1i$ than for ${\cal E}h1w$, as can be inferred
from comparing the potential energies at $t=0.0$. Although still
supercritical, the field in model ${\cal E}h1i$ is already strong
enough to suppress motions perpendicular to the mean field, so that
the field strength perpendicular to its initial mean direction is
small, while strong shocks/waves can be formed parallel to the mean field,
as observed as well by Smith et al. (2000).
Thus, somewhat surprisingly, the density enhancements are larger for
the stronger-field model ${\cal E}h1i$ than for the weak-field model
${\cal E}h1w$, leading to earlier collapse, as seen in
Figure~\ref{fig:kmodes}.  In model ${\cal E}h1w$ the weaker, more
turbulent magnetic field produces a more isotropic magnetic pressure
that cushions and broadens the shocks, thus decreasing the density
enhancements and delaying collapse.

\begin{figure}
\epsfig{file=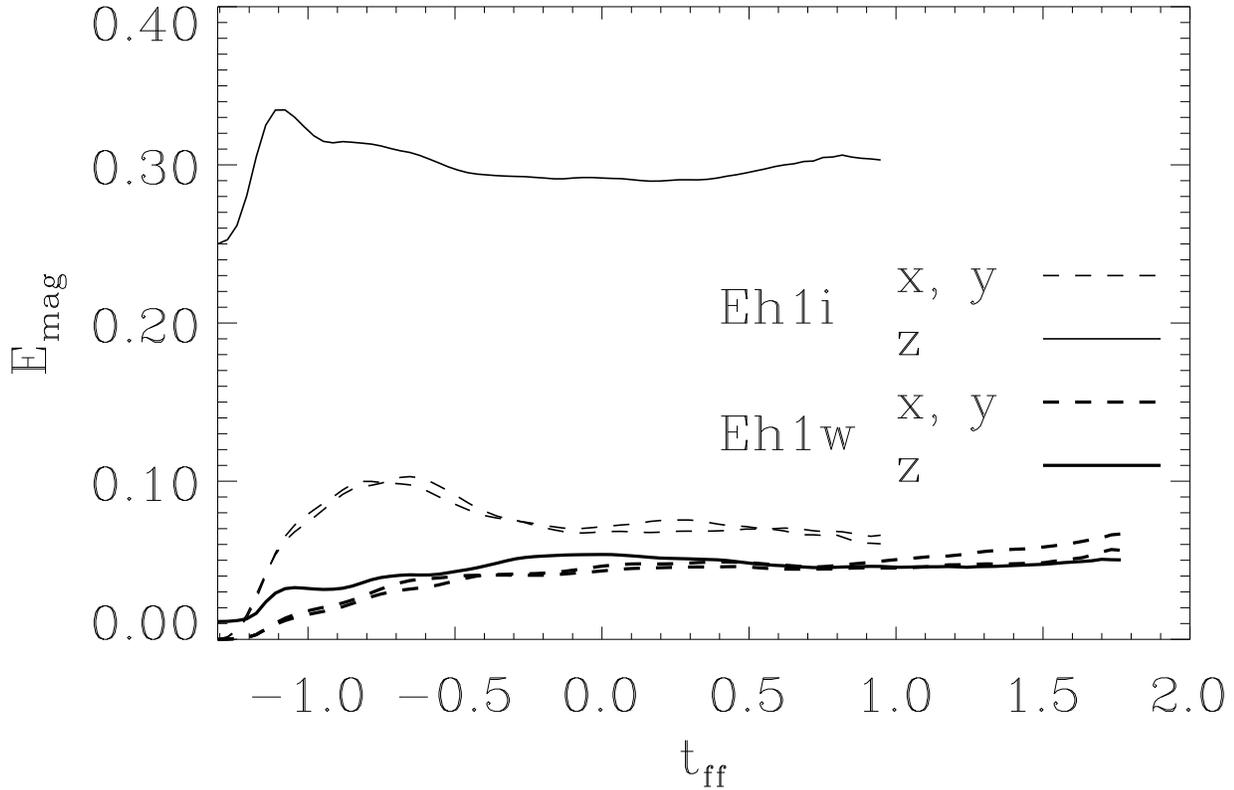,height=16.0cm,angle=90}
\caption{\label{fig:mageng-comp}
Magnetic energy components for magnetic models ${\cal E}h1w$ 
(thick lines) and ${\cal E}h1i$ (thin lines) against time. 
Solid lines denote the energy of the $z$-component (initial mean field
orientation), dashed lines the $x$- and $y$-components. Gravity is
turned on at $t=0.0$. In the weakly magnetized model ${\cal E}h1w$, the 
field has ceased to show a net orientation at $t=0.0$. It has developped
a fully turbulent state. The strong field (${\cal E}h1i$) continues to
keep its mean orientation.}
\end{figure}

We illustrate this effect in Figure~\ref{fig:mageng-comp}, where we
plot the $x$- $y$- and $z$-components of the magnetic energy against
time for the magnetized models ${\cal E}h1w$ and ${\cal E}h1i$.  In
the weak-field case, the field is quickly tangled by the flow, so that
it has no preferred direction by $t=0.0$, when gravity is switched on.
The magnetic energy, and so the magnetic pressure, is isotropic.
There is no secular increase of any component with time, supporting
the picture of local collapse: In a globally collapsing environment,
the magnetic field lines would follow the global gas flow and lead to
a noticeable increase of magnetic energy.  In the stronger-field case,
on the other hand, the flow is dominated by the mean field oriented
along the $z$-axis. The field allows matter to move more freely in the
parallel than in the perpendicular direction.  Matter thus collapses
preferentially along field lines first, and then globally collapses.

After one free-fall time all the models have collapsed
(Figure~\ref{fig:engspec-t=0}). Note that, as in
Paper I, $E_{pot}(k) > E_{kin}(k)$ for all $k$ does not necessarily
mean that the model becomes globally unstable. With increasing time,
$E_{pot}(k)$ becomes constant for all $k$ just because this is the
Fourier transform of a $\delta$-function, signifying that local
collapse has produced point-like high-density cores. 
A similar argument applies for the kinetic energy: The flat spectrum
stems from local concentrations of kinetic energy around the
collapsing regions. Here, the spectral analysis no longer yields 
information on global stability.
 
\subsubsection{Conclusions\label{subsubsec:discussion}}

We conclude that the delay of local collapse seen in our magnetized
simulations is caused mainly by weakly magnetized turbulence acting as
a more or less isotropic pressure in the gas, decreasing density
enhancements due to shock interactions. We feel justified in claiming
that magnetic fields, as long as they do not provide magnetostatic
support, can {\em not} prevent local collapse, even in the presence of
supersonic turbulence.

We note that once bound cores form, we take this as evidence for local
collapse, although subsequent shock interactions may destroy these
cores again.  In a real cloud, ambipolar diffusion would set in at the
length scale of transient cores, so that any internal turbulence would
be quickly dissipated, allowing further collapse, as discussed in Paper~I.

A large-scale driver, such as interacting supernova remnants or
galactic shear, together with magnetic fields, seems to act like a
driver with a smaller effective scale in the sense that both yield a
more uniform core distribution and a somewhat slower collapse rate.
In weakly magnetized turbulence, a more or less isotropic
magnetic pressure reduces the density enhancements behind shocks and
thus slows down the process of isolated collapse. In strongly
magnetized turbulence, however, the mean magnetic field dominates. The
magnetic pressure is not isotropic any more, so the shocks
perpendicular to the mean field direction cause high enough
density enhancements for the regions to collapse within a free-fall
time.

Thus, for small field strength, the effective additional pressure
may be represented by a simple pressure term. However, in the 
regime of field strength interesting for molecular clouds, the field,
although supercritical, is strong enough to result in an anisotropic
magnetic pressure.
Magnetic turbulence is an all-scale non-isotropic phenomenon, and the
compression and perturbations on large scales make the cloud finally
collapse. 

\section{Summary \label{sec:conclusions}}

In this paper, we investigated whether magnetized turbulence can prevent
collapse of a Jeans-unstable region. From our high-resolution simulations
we conclude that
\begin{enumerate}
  \item In order to resolve self-gravitating MHD turbulence using a
        grid-based method like ZEUS-3D, the local Jeans length should
        not fall short of at least four grid zones for magnetostatic support
        and six grid zones for magnetodynamic support.
  \item Local collapse cannot be prevented by
        magnetized turbulence in the absence of mean-field
        support. Strong local density enhancements due to shock
        interactions start collapsing at once. 
  \item However, the magnetic fields do delay local collapse by 
        decreasing local density enhancements via magnetic pressure behind shocks.
  \item Weakly magnetized turbulence appears qualitatively similar
        to hydrodynamic turbulence driven on a slightly smaller scale,
        while stronger fields close to but under the value for
        magnetostatic support tend to organize the flow into sheets
        and allow more clustered collapse.
  \item The strength and wavelength of turbulent driving governs the
        behaviour of the cloud, overshadowing the effects of
        magnetic fields that do not provide magnetostatic support. 
  \item MHD turbulence can not prevent local collapse for much longer than 
        a global free-fall time. Stars begin to form at a low rate as soon 
        as local density
        enhancements contract. This result favours a dynamical picture
        of molecular clouds being a transient feature in the ISM
        (Ballesteros-Paredes et al. 1999, Elmegreen 2000) rather than
        living for many free-fall times.
\end{enumerate}

\acknowledgments We thank A.~Burkert, H.~Lesch, S.~Phleps, E.~V\'{a}zquez-Semadeni 
and E.~Zweibel for valuable discussions. Thanks go to the referee, whose criticism
helped clarify important points in the paper. 
M-MML is partially supported by an NSF CAREER fellowship, grant number
AST 99-85392. Computations presented here 
were performed on SGI Origin 2000 machines of the Rechenzentrum Garching of 
the Max-Planck-Gesellschaft, the National Center for Supercomputing 
Applications (NCSA), and the Hayden Planetarium.  ZEUS was used by courtesy 
of the Laboratory for Computational Astrophysics at the NCSA. This research 
has made use of
NASA's Astrophysics Data System Abstract Service.

\clearpage

\begin{table}
\begin{tabular}{ccccccccc}
\hline
Name         &Resolution
             &$k_{drv}$
             &$M_{J}^{turb}$
             &$\beta$
             & $\beta_{turb}$
             &$M/M_{cr}$ 
             &$t_{5\mbox{\%}}$ \\
\hline
${\cal D}h1$ & $256^3$   & $1-2$  & $15$ &$\infty$&$\infty$&$\infty$&$0.4$   \\
${\cal D}h3$ & $256^3$   & $7-8$  & $15$ &$\infty$&$\infty$&$\infty$&$0.7$   \\
${\cal E}h1w$& $256^3$   & $1-2$  & $15$ &$0.9$   &$87.2$  &$8.3$   &$1.4$   \\
${\cal E}i1w$& $128^3$   & $1-2$  & $15$ &$0.9$   &$87.2$  &$8.3$   &$1.3$   \\
${\cal E}l1w$& $64^3$    & $1-2$  & $15$ &$0.9$   &$87.2$  &$8.3$   &$0.6$   \\
${\cal E}h1i$& $256^3$   & $1-2$  & $15$ &$0.04$  &$3.9$   &$1.8$   &$0.8$   \\
${\cal E}i1i$& $128^3$   & $1-2$  & $15$ &$0.04$  &$3.9$   &$1.8$   &$0.4$   \\
${\cal E}l1i$& $64^3$    & $1-2$  & $15$ &$0.04$  &$3.9$   &$1.8$   &   \\
${\cal E}i1s$& $128^3$   & $1-2$  & $15$ &$0.01$  &$1.0$   &$0.8$   &$1.5$   \\
${\cal E}l1s$& $64^3$    & $1-2$  & $15$ &$0.01$  &$1.0$   &$0.8$   &$1.0$   \\
${\cal E}i0s$& $128^3$   & $0$    & $-$  &$0.01$  &$-$     &$0.8$   &$-$     \\
${\cal E}l0s$& $64^3$    & $0$    & $-$  &$0.01$  &$-$     &$0.8$   &$-$     \\
${\cal F}l0w$& $64^3$ & $0$ & $-$ &$3\cdot 10^{-3}-0.14$ &$-$&$1.10$   &$-$  \\
${\cal F}l0i$& $64^3$ & $0$ & $-$ &$2\cdot 10^{-3}-0.09$ &$-$&$0.88$   &$-$  \\
${\cal F}l0m$& $64^3$ & $0$ & $-$ &$1\cdot 10^{-3}-0.04$ &$-$&$0.59$   &$-$  \\
${\cal F}l0s$& $64^3$ & $0$ & $-$ &$3\cdot 10^{-4}-0.02$ &$-$&$0.44$   &$-$  \\
${\cal G}i1$ & $128^3$ & $1-2$& $19$ &$\infty$&$\infty$&$\infty$&$1.2$  \\
${\cal G}i1w$& $128^3$ & $1-2$& $19$ &$4.04$ &$100.0$ &$8.3$   &$2.1$   \\
${\cal G}i1i$& $128^3$ & $1-2$& $19$ &$0.23$ &$5.9$   &$2.0$   &$1.0$   \\
${\cal G}i1m$& $128^3$ & $1-2$& $19$ &$0.07$ &$1.8$   &$1.1$   &$(3.5)$ \\
${\cal G}i1s$& $128^3$ & $1-2$& $19$ &$0.05$ &$1.1$   &$0.8$   &$-$     \\
\hline
\end{tabular}
\caption{Parameters of models used. $M_J^{turb}=\rho^{1/2}(\pi/G)^{3/2}
(c_s^2+\langle v^2\rangle /3)^{3/2}$ gives the turbulent Jeans mass,
$\beta=P_{th}/P_{mag}=8\pi c_s^2 \rho/B^2$ and
$\beta_{turb}=P_{turb}/P_{mag}$. $t_{5\mbox{\%}}$ denotes the time at
which $5$\% of the total mass has been accreted onto cores. Times have
been normalized to the global free-fall time.  $M/M_{cr}$ gives the
ratio of cloud mass to critical mass according to equation
\ref{equ:magstatsup}. Models ${\cal E}l1i$ and ${\cal E}l1s$ have been
computed three times with varying random seeds to check the influence
of the velocity field on the results. They are not listed explicitly.
Models ${\cal E}l0s$, ${\cal E}i0s$ and ${\cal F}$ are non-driven runs, used for
the analysis of numerical diffusion. Model ${\cal G}i1s$ collapses so
slowly, that it only reached $M_*=3$\% after $t=3.5t_{\rm ff}$.
\label{tab:models}}
\end{table}

\begin{table}
\begin{tabular}{cccccccc}
\hline
model        &$c_s$
             &$n_J$
             &${\cal M}$
             & $\lambda^{pk}_J$
             &$64^3$ &$128^3$&$256^3$ \\
\hline
${\cal D/E}$ &$0.10$  &$64$      & $10$     & $0.05$     &$1.6$&$3.2$&$6.4$ \\
${\cal G}$   &$0.213$ &$6.4$     & $5$      & $0.11$     &$3.4$&$6.8$&$13.6$\\
\hline
\end{tabular}
\caption{Peak Jeans lengths for all turbulent models, where
         $\lambda^{pk}_J$ is determined via the peak density
         $\rho_{pk}= {\cal M}^2 \rho_0$. The second column lists the sound 
         speed $c_s$, followed by the Mach number ${\cal M}$ and the number
         of thermal Jeans masses in the box $n_J$.
         The last three columns contain the Jeans lengths in zones for the resolution 
         denoted in the top row. 
         \label{tab:lambda-j}}
\end{table}

\clearpage

\end{document}